\documentclass[doublecol]{epl2} 
\usepackage{subfigure}

\title{Local measurement of vortex statistics in quantum turbulence}

\author{Eric Woillez, Jérôme Valentin  \and Philippe-E. Roche}

\institute{                    
    \inst{} Univ. Grenoble Alpes, CNRS, Institut NEEL, F-38042 Grenoble, France
}

\abstract{
The density fluctuations of quantum vortex lines are measured in a turbulent flow of superfluid He, at temperatures corresponding to superfluid fraction of 16\%, 47\% and 81\%. The probe is a micro-fabricated second sound resonator that allows for local and small-scale measurements in the core of the flow, at a 10-mesh-size behind a grid. Remarkably, all the vortex power spectra collapse on a single master curve, independently from the superfluid fraction and the mean velocity. By contrast with previous measurements, we report an peculiar shape of the power spectra. The vortex density probability distributions are found to be strongly skewed, similarly to the vorticity distributions observed in classical turbulence. Implications of those results are discussed. 
}

\begin{document}

\maketitle

\section{Introduction}
In the zero temperature limit, quantum fluids behave at the macroscopic scale as a single coherent quantum state, the superfluid \cite{DonnellyLivreVortices}.  Compared to classical fluids, the quantum coherence of superfluids creates a strong additional constraint on the velocity field, namely to be irrotational. Rotational motion can only appear when the macroscopic coherence of the wave function is broken by topological defects called quantum vortices. In that case, the circulation of the velocity around the quantum vortex has a fixed value ($\kappa \simeq 10^{-7}$m$^2$s$^{-1}$ in $^4$He). Turbulence in superfluids can be thought of as an intricate process of distortion, reconnection and breaking of those topological singularities \cite{BarenghiSkrbekSreenivasan_IntroPNAS2014}, but in such a way that the system seems to mimic the classical turbulence at large scales \cite{spectra:PNAS2014}. This has been particularly obvious in the velocity spectra probed with a variety of anemometers, in highly turbulent flows \cite{Maurer1998,salort2010turbulent,Salort:EPL2012,rusaouen2017intermittency} or in the measurement of vortex bundles using parietal pressure probes \cite{Rusaouen:parietalEPL2017}. In some sense, quantum turbulence is an irreducible model, or to say it differently, is a kind of "skeleton" for all types of turbulence. 

At finite temperature, the quantum fluid is not a pure superfluid: it behaves as if it experienced friction with a background viscous fluid, called the ``normal fluid''. The relative mass density of the superfluid $\rho_s/\rho$ (where $\rho$ is the total mass density) decreases from one at 0~K to zero at the superfluid transition temperature ($T_\lambda \simeq 2.18$~K in $^4$He). The presence of a finite normal fluid fraction allows for propagation of temperature waves - a property referred to as ``second sound"- which opens the rare opportunity to probe directly the presence of the quantum vortices \cite{DonnellyPhysicsToday1984}. 

This is done in the present article, where the statistics of superfluid vortex lines density $\mathcal{L}$ are locally measured by ``second sound tweezer" (see the description in paragraph``probes"), over one and a half decade of the inertial scales, and over a wide range of $\rho_s/\rho$ spanning from 0.16 to 0.81. Surprisingly, the result does not corroborate the widespread idea that the large scales of quantum turbulence reproduce those of classical turbulence: the measured  spectra of $\mathcal{L}$ (see Fig. \ref{fig:spectres})  differs from classical-like enstrophy spectra \cite{baudet1996spatial,ishihara2003spectra}. Besides, it also differs from  the only\footnote{
Literature also reports experimental \cite{Bradley:PRL2008} and numerical \cite{FujiyamaJLTP2010, BaggaleyPRB2011,Baggaley_VLD:PRL2012,BaggaleyPRL2015,tsepelin2017visualization}  spectra of the vortex line density spatially integrated across the whole flow. Still, spectra of such ``integral'' quantities differ in nature from the spectra of local quantities, due to strong filtering effects of spatial fluctuations.}
previous direct measurement of $\mathcal{L}$ with second sound tweezers \cite{roche2007vortex} at $\rho_s/\rho \simeq 0.84$. 

The measurement of the vortex lines density provides one of the very few constraints for the disputed modeling of the small scales of quantum turbulence. Even after intense numerical \cite{salort2011mesoscale,Baggaley_Coherentvortexstructures_EPL2012}  and theoretical \cite{RocheInterpretation:EPL2008,Nemirovskii:PRB2012,boue2015energyVorticity} studies,  the statistics of quantum vortices show that even the large scales of quantum flows can still be surprising.

\section{Experimental setup}

\begin{figure}
\begin{centering}
\includegraphics[height=10cm]{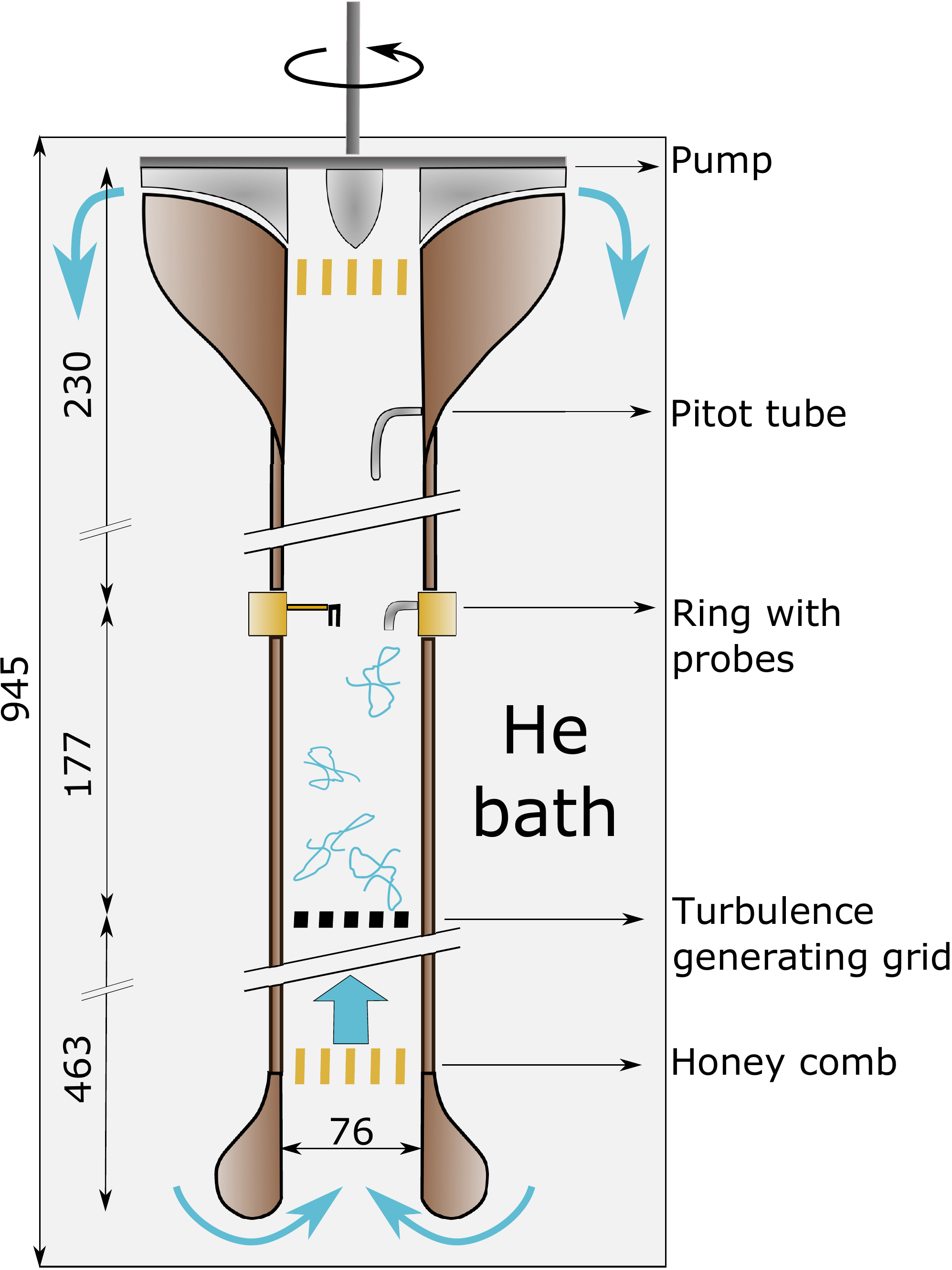}
\par\end{centering}
\caption{Sketch of the flow and the experimental setup with
probes. \label{fig:schema_toupie}}
\end{figure}

The experimental setup has been described in details in a previous publication
\cite{rusaouen2017intermittency},
and we only review in this section the major modifications. The setup
consists in a wind tunnel inside a cylindrical cryostat (see
Fig. \ref{fig:schema_toupie}) filled with He-II.
The flow is continuously
powered by a centrifugal pump located at the top of the tunnel. At
the bottom, an optimized 3D-printed conditioner ensures a smooth
entry of the fluid, without boundary layer detachment, inside a pipe of $\Phi=76$ mm inner diameter. Spin motion is broken by radial screens built in the conditioner. The fluid
is then ``cleaned'' again by a 5-cm-long and $3$-mm-cell honeycomb. The mean flow velocity $U$ is measured with a Pitot tube located $130$ mm upstream the pipe outlet. We allow a maximal mean velocity $U=1.3$ m/s inside the pipe to avoid any cavitation effect with the pump.

The main new element compared to the previous design  
is a mono-planar grid located $177$ mm upstream the probes to generate
turbulence. The grid has a $M=17$ mm mesh with square bars of thickness $b=4$ mm, which gives a porosity of $\beta=(1-b/M)^{2}\approx0.58$. 

The choice to position the probes at a distance $\sim 10M$ downstream the grid is the result of a compromise between the desire to have a ``large'' turbulence intensity, and the necessity to leave enough space for turbulence to develop between the grid and the probes. According to \cite{vita2018generating}, this distance is enough to avoid near-field effects of the grid. However, we emphasize that our main experimental results (Fig. \ref{fig:spectres}-\ref{fig:histogrammes}) do not depend on perfect turbulent isotropy and homogeneity. In-situ measurements of the mean vortex line density can be used to indirectly (via Eq. \ref{eq:tau}) give an estimation of the turbulence intensity $\tau=u^\mathrm{rms}/U \simeq 12-13\%$ (where $u^\mathrm{rms}$ is the standard deviation of longitudinal velocity component). We present the results later in Fig. \ref{fig:scalingU}.
For comparison, Vita and co. \cite{vita2018generating} report a turbulence intensity around $\tau=9\%$ percents at $10M$ in a classical grid flow of similar porosity. The difference between both values of $\tau$ could originate from a prefactor uncertainty in Eq. (\ref{eq:tau}) or from differences in flow design (e.g. the absence of a contraction behind the honeycomb). This difference has no important consequences for the measurement of quantum vortex statistics.

The longitudinal integral length scale of the flow $H\simeq 5.0$~mm is assessed by fitting velocity spectra (see bottom panel of Fig. \ref{fig:spectres}) with the von K\'arm\'an formula (eg. see \cite{vita2018generating}). For comparison, the integral scale reported for the similar grid in \cite{vita2018generating}, once rescaled by the grid size, gives a nearby estimate of $7.4$ mm.

The Reynolds number $Re$ defined with $u^\mathrm{rms} H$ and the kinematic
viscosity $1.8\times10^{-8}$~m$^2$s$^{-1}$ of liquid He just above $T_\lambda$, is $Re=3.3\times10^4$ for $U=1$ m/s. Using standard homogeneous isotropic turbulence formula, the Taylor scale Reynolds number is $R_\lambda=\sqrt{15Re}\approx 700$ (for $\tau=12\%$ and $H=5$ mm). This gives an indication of turbulence intensity of the flow below $T_\lambda$.

Temperature of the helium bath is set via pressure regulation gates.
The exceptional thermal conductivity of He-II ensures an homogeneous
temperature inside the bath for $T<T_{\lambda}$. Two  Cernox
thermometers, one located just above the pump, the other one on the
side of the pipe close to the probes, allow for direct monitoring
of $T$.

\section{Probes}

\begin{figure}
\begin{centering}
\includegraphics[height=6cm]{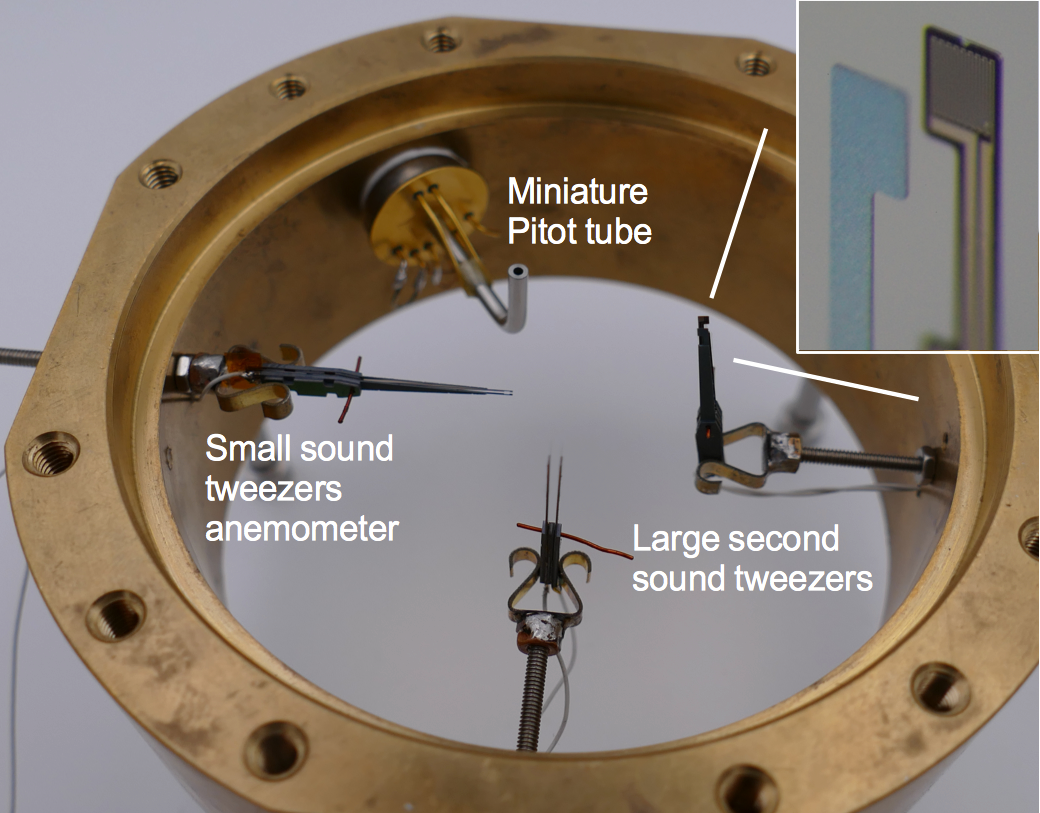}
\par\end{centering}
\caption{Ring with probes. The inset is a zoom on the heating and the thermometer plates of a second sound tweezers. The Pitot tube is not used in the present experiment. \label{fig:probes}}
\end{figure}

Our probes are micro-fabricated second sound tweezers
of the millimeter size according to the same principle as in \cite{roche2007vortex}.
As displayed in the inset of Fig. \ref{fig:probes}, the tweezers are composed
of one heating plate and one thermometer plate facing each other and
thus creating a resonant cavity for thermal waves. The heating
plate generates a stationary thermal wave of the order of $0.1$
mK between the plates, the amplitude of which can be recorded by the
thermometer plate. Two major improvements have been done compared
to the tweezers in \cite{roche2007vortex} : first, the length of
the arms supporting the plates has been increased to \textbf{$14$}
mm to avoid blockage effects due to the stack of silicon wafers (about 1.5 mm thick) downstream the cavity. Second,
two notches are done in the arms to avoid interference due to additional
reflections of the thermal wave on the arms. Further details will be given in a future publication. 

In the presence of He flow, a variation of the amplitude and phase
of the thermal wave can be observed. This variation is due 
to two main physical effects. The presence of quantum vortex lines
inside the cavity causes an attenuation of the wave \cite{DonnellyPhysicsToday1984,varga2019} with a very
minor phase shift \cite{miller1978velocity}. This attenuation can be very accurately modelized by
a bulk dissipation coefficient inside the cavity denoted $\xi_{L}$. The second effect is a ballistic advection
of the wave out of the cavity. It is related to both an attenuation of
the temperature oscillation and an important phase shift. Depending
on the flow mean velocity $U$, the size of the tweezers, and the
frequency of the wave, one of these two effects can overwhelm the
other. We have thus designed two models of tweezers: one model to
take advantage of the first effect to measure the vortex lines density (VLD), and the other one 
to take advantage of the second effect to measure the velocity.

The two largest tweezers displayed in Fig. \ref{fig:probes} are designed to measure the quantum vortex
lines density. The plates size is $l=1$ mm and the gaps
between the plates are $D=1.32$ mm and $D=0.83$ mm respectively.
The plates face each other with positioning accuracy of a few micrometers.
The tweezers are oriented parallel to the flow (see Fig. \ref{fig:probes}, the mean flow is directed from top to bottom)
to minimize the effect of ballistic advection of the wave. 

The smallest tweezers displayed in Fig. \ref{fig:probes} are designed to be mainly sensitive to the velocity fluctuations
parallel to the mean flow. The two plates have a size $l=250$ $\mu$m,
and are separated by a gap of $D=0.431$ mm. The tweezers are oriented
perpendicular to the mean flow (see Fig. \ref{fig:probes})
with an intentional lateral shift of the heater and the thermometer
of about $l/2$. This configuration is expected to maximize the sensitivity
to ballistic advection, and thus to velocity fluctuations. To second order however, the probe still keeps sensitivity to the quantum vortices produced both by turbulence and by the intense heating of the plates, that's why we were not able to calibrate it reliably. The (uncalibrated) spectrum of this probe (see bottom panel of Fig. \ref{fig:spectres}) is only used to estimate the integral length scale. The role of this probe is also to prove that the signal statistics of the largest tweezers are not due to velocity fluctuations.

\section{Method}
\begin{figure}
\begin{centering}
\includegraphics[height=4.5cm]{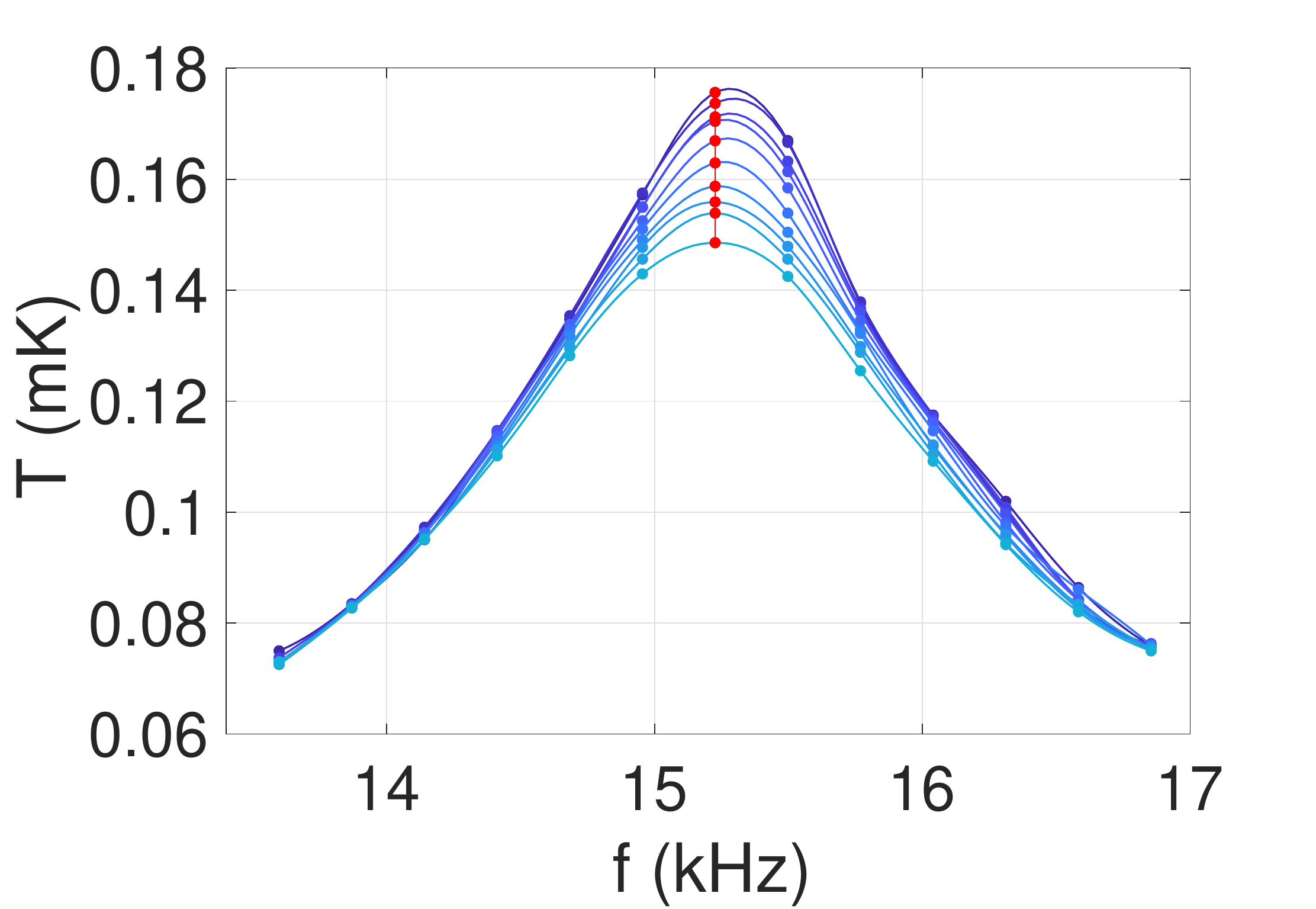}\\
\includegraphics[height=5cm]{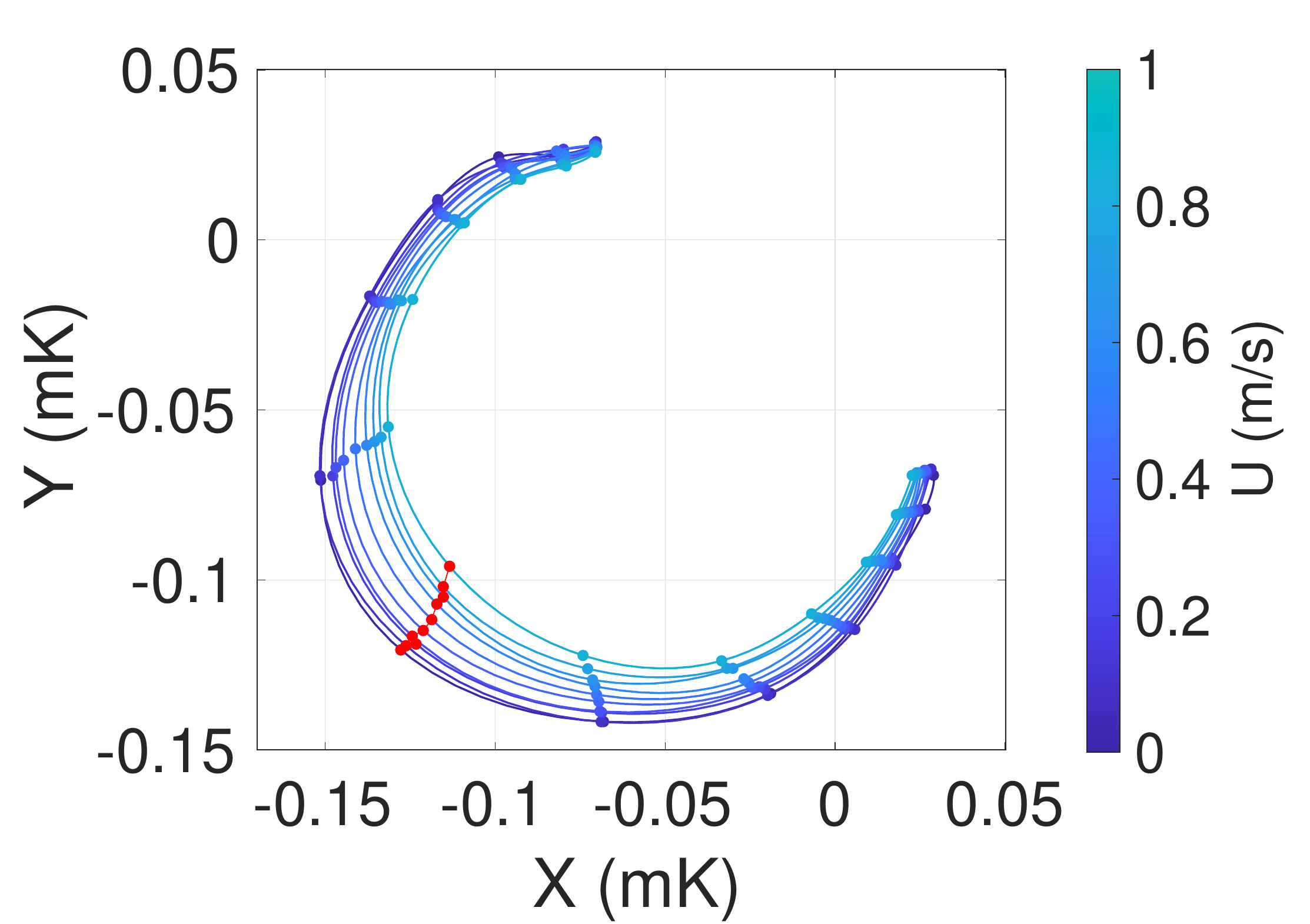}
\par\end{centering}
\caption{\textbf{Top: }second sound resonance of one of the large tweezers around $15.2$
kHz. The value of $U$ increases from top curve to bottom curve. The vertical axis gives the amplitude
of the thermal wave in K. \textbf{Bottom:} representation of the
same resonance in phase and quadrature.\label{fig:methode}}
\end{figure}

Figure \ref{fig:methode} displays a resonance of a large tweezers
at frequency $f_{0}=15.2$ kHz, for increasing values of the mean velocity. The temperature oscillation $T$ measured by the thermometer
is demodulated by a Lock-in amplifier NF LI5640. $T$ can be accurately fitted by a classical
Fabry-Perot formula
\begin{equation}
T=\frac{A}{\sinh\left(i\frac{2\pi(f-f_{0})D}{c_{2}}+\xi D\right)} \label{eq:FP formula}
\end{equation}
where $i^2=-1$, $f_{0}$ is the resonant frequency for which the wave locally reaches its maximal
amplitude, $c_{2}$ is the second sound velocity, $A$ is a parameter
to be fitted, and $\xi$ is related to the energy loss of
the wave in the cavity. The top panel of Fig. \ref{fig:methode}
displays the amplitude of the thermal wave (in mK) as a function
of the frequency, and the bottom panel shows the same signal in phase and quadrature. When the
frequency is swept, the signal follows a curve close to a circle crossing
the point of coordinates $(0,0)$. Fig. \ref{fig:methode} clearly shows that the
resonant peak shrinks more and more when $U$ increases, which is
interpreted as attenuation of the wave inside the cavity. The red points
display the attenuation of the signal at constant value of $f$. It can
be seen on the bottom panel that the variation of the signal is close
to a pure attenuation, that is, without phase shift. 
$\xi$ can be decomposed as
\begin{equation}
\xi=\xi_{0}+\xi_{L}\label{eq:xi_dec}
\end{equation}
 where $\xi_{0}$ is the attenuation factor when $U=0$ m/s and $\xi_{L}$
is the additional attenuation created by the presence of quantum vortex
lines inside the cavity. $\xi_{L}$ is the signal of interest as
it can be directly related to the vortex lines density (VLD) using
the relation
\begin{eqnarray}
\xi_{L} & = & \frac{B\kappa L_{\perp}}{4c_{2}},\label{eq:VLD}\\
L_{\perp} & = & \frac{1}{\mathcal{V}}\int\sin^{2}\theta(l){\rm d}l\label{eq:VLD def}
\end{eqnarray}
where $B$ is the first Vinen coefficient, $\kappa\approx9.98\times10^{-8}$
m$^{2}$/s is the quantum of circulation, $\mathcal{V}$ is the cavity
volume, $l$ is the curvilinear absciss along the vortex line, $\theta(l)$
is the angle between the vector tangent to the line and the direction
perpendicular to the plates. We note that the summation is weighted by the distribution of the second sound nodes and antinodes inside the cavity and does not exactly corresponds to a uniform average but we neglect this effect in the following. Our aim is to measure both the average value
and the fluctuations of $L_{\perp}$, as a function of $U$ and the superfluid fraction.

The method goes as follows: first, we choose a resonant frequency
$f_{0}$ where the amplitude of the signal has a local maximum and
we fix the frequency of the heating to this value $f_{0}$. Then we
vary the mean velocity $U$ and we record the response of the thermometer
plate in phase and quadrature. The measurements
show that the velocity-induced displacement in the complex plane follows a straight line in a direction
$\overrightarrow{e}$ approximately orthogonal to the resonant curve.
Expressions (\ref{eq:FP formula}-\ref{eq:xi_dec}) give $\xi_{L}$
from the measured amplitude $T$ by\cite{roche2007vortex}
\begin{equation}
\xi_{L}=\frac{1}{D}\rm{asinh}\left(\frac{A}{T}\right)-\xi_{0}.\label{eq:displacement}
\end{equation}

\begin{figure}
\begin{centering}
\includegraphics[height=7cm]{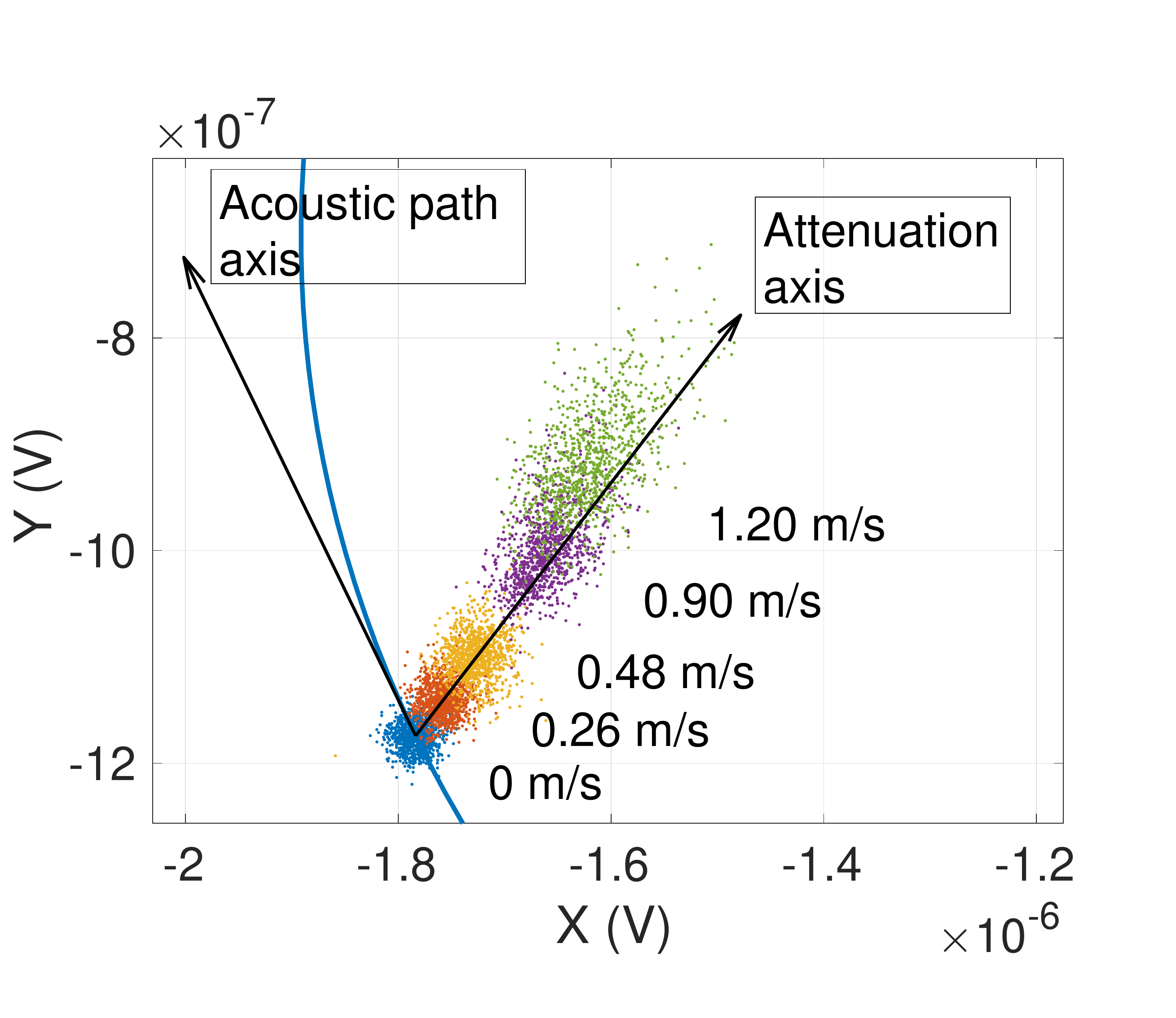}
\par\end{centering}
\caption{Fluctuations of the thermal wave in phase and quadrature. The colored
clouds show the fluctuations of the signal, for different values of
$U$. The blue curve shows the resonance for $U=0$ m/s. The fluctuations tangent
to the resonant curve are created by a variation of the acoustic path.
The quantum vortices are associated to attenuation of the wave and create
a displacement along the attenuation axis. \label{fig:fluctuations}}
\end{figure}

The colored dots of Fig. \ref{fig:fluctuations} illustrate the fluctuations of the signal
in phase and quadrature, for different values of $U$. The average signal
moves in the direction of the attenuation axis. The figure also shows
a part of the resonant curve for $U=0$. The fluctuations have two
components in the plane, both associated with different physical
phenomena. Fluctuations in the direction tangent to the resonant curve
can be interpreted as a variation of the acoustic path $\frac{2\pi(f-f_{0})D}{c_{2}}$
without attenuation of the wave. Those fluctuations can occur for example
because the two arms of the tweezers vibrate with submicron amplitude,
or because the temperature variations modify the second sound velocity
$c_{2}$. To isolate only the fluctuations associated to attenuation by
the quantum vortices, we split the signal into a component along the attenuation axis, and another one along the acoustic path axis.
We then convert the displacement along the attenuation axis into vortex line density (VLD) using
expressions (\ref{fig:methode}-\ref{eq:displacement}).

\section{Results}

As a check of the validity of our approach, we measured the average
response of the second sound tweezers as a function of the mean velocity
$U$. According to literature\cite{Babuin:EPL2014}, we were expecting the scaling $\left\langle L_{\perp}\right\rangle^2 \propto U^{3}$, with a prefactor related to the flow main characteristics.  The function $\left\langle L_{\perp}\right\rangle $ was thus
measured for a range $0.4<U<1.25$ m/s with a time averaging over
$300$ ms, at the three different temperatures $1.65$ K, $1.99$ K and $2.14$ K. 

\begin{figure}
\begin{centering}
\includegraphics[height=6cm]{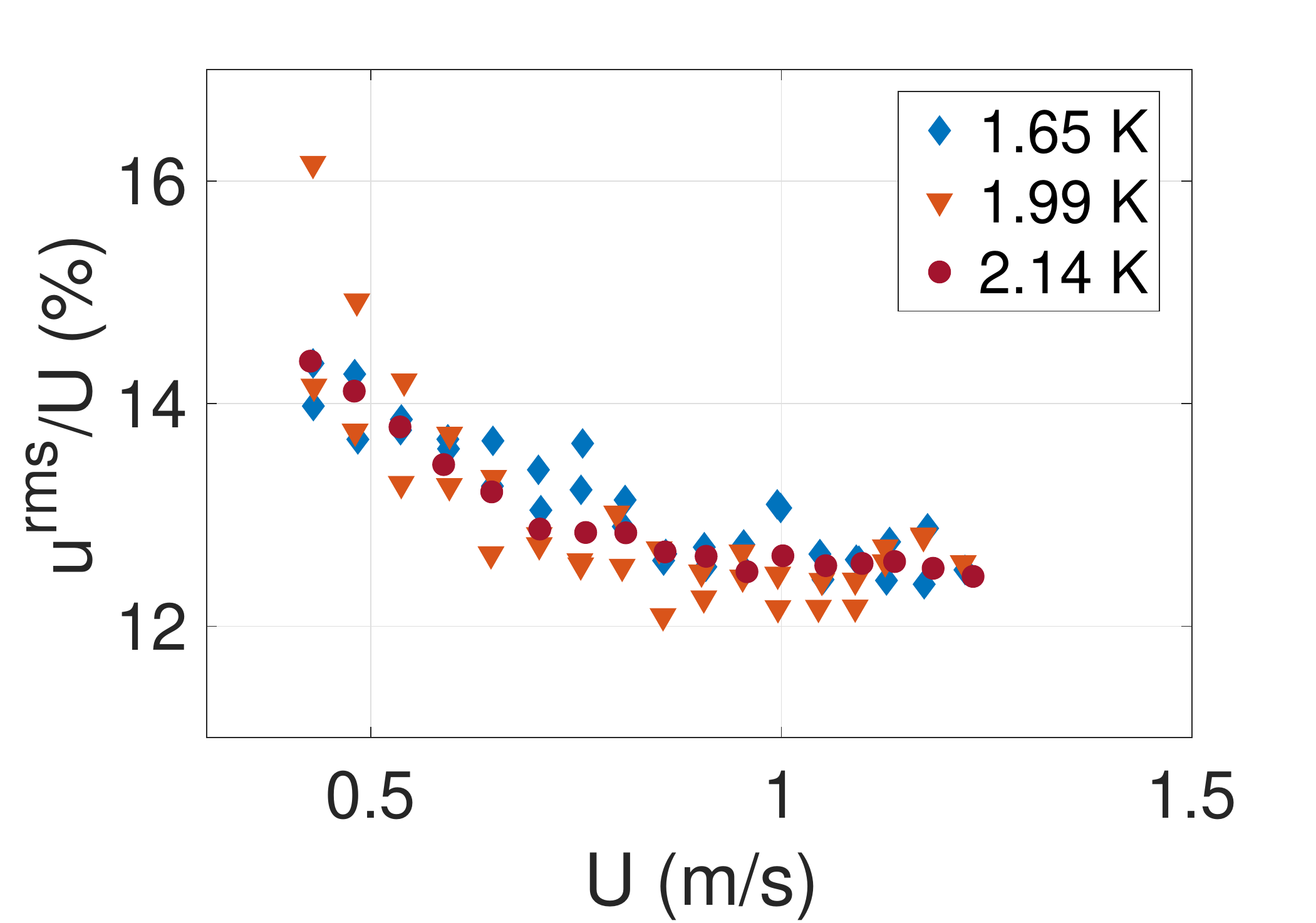}
\caption{Indirect measurement of the turbulence intensity $\tau=u^{\rm{rms}}/U$ as a function of $U$ using Eq. (\ref{eq:tau}). The three different symbols correspond to three values of the mean temperature.   \label{fig:scalingU}}
\par\end{centering}
\end{figure}

An effective superfluid viscosity $\nu_\mathrm{eff}$ is customarily defined in quantum turbulence by $\epsilon = \nu_\mathrm{eff} ( \kappa \mathcal{L} )^2$ where $\epsilon$ is the dissipation and $\mathcal{L}=3\left\langle L_{\perp}\right\rangle/2$ is the averaged VLD (we assume isotropy of the tangle)\cite{Vinen:JLTP2002}. For large $R_\lambda$ homogeneous isotropic flows, we also have $\epsilon \simeq 0.79\, U^3\tau^3/H$ (eg see \cite{pope:book} p.245), which entails
\begin{equation}
    \tau^3 \simeq 2.85  \frac{ \nu_\mathrm{eff} H \kappa^2\left\langle L_{\perp}\right\rangle^2}{U^3} \label{eq:tau}
\end{equation}
Using Eq. (\ref{eq:tau}), we compute the turbulence intensity as a function of $U$, for the three considered temperatures. The result is displayed in Fig. \ref{fig:scalingU}. The figure shows that the turbulence intensity reaches a plateau of about $12\%$ above $0.8$ m/s, a value in accordance with the turbulence intensity of $9\%$ reported in \cite{vita2018generating} for a grid turbulence with similar characteristics. The figure also confirms that the expected scaling $\left\langle L_{\perp}\right\rangle^2 \propto U^{3}$ is reached in our experiment for the range of velocities $U>0.8$ m/s.  

The temperature-dependent viscosity $\nu_\mathrm{eff}$ in Eq. (\ref{eq:tau}) has been measured in a number of experiments (e.g. see compilations in \cite{Babuin:EPL2014,boue2015energyVorticity,gao2018dissipation}). Still, the uncertainty on its value exceeds a factor 2. For the temperatures $1.65$ K and $1.99$ K, we used the average values $0.2 \kappa$ and $0.25\kappa$. By lack of reference experimental value of $\nu_\mathrm{eff}$ above $2.1$ K, we determined it by collapsing the $\tau (U)$ datasets obtained at $2.14$ K with the two others. We found the value $\nu_\mathrm{eff}\approx 0.5\kappa$ at $2.14$ K.

 Assuming isotropy of the vortex tangle, the value of $\mathcal{L} $ gives
a direct order of magnitude of the inter-vortex spacing $\delta=1/\sqrt{\mathcal{L}}$. We find $\delta\approx5$
$\mu m$ at 1.65 K and a mean velocity of 1 m/s. This shows the large scale separation between the inter-vortex spacing and the flow integral scale $H$, a confirmation of an intense turbulent regime.

\begin{figure}
\begin{centering}
\includegraphics[height=9cm]{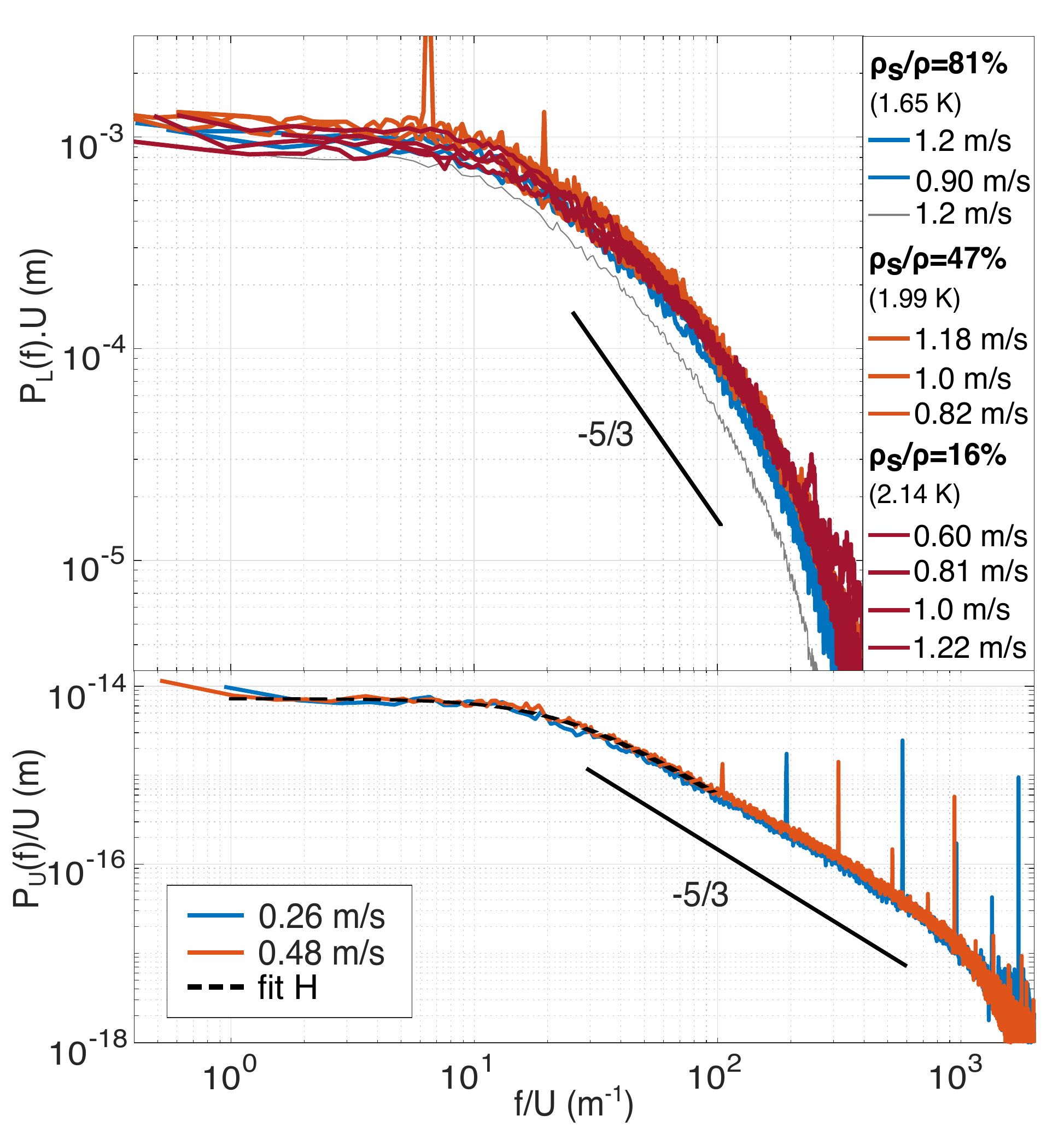}\\
\par\end{centering}
\caption{\textbf{Top:} Power spectral density of the projected vortex line density (VLD) $L_{\perp}$, obtained
with the large second sound tweezers, for different values of $U$
and temperatures. All measured spectra collapse using the scaling
$f/U$ and $P_L(f)\times U$. The fluctuations have been rescaled by the mean
value of the VLD such that the integral of the above curves directly
give the VLD turbulence intensity.
\textbf{Bottom:} Power spectral density of the uncalibrated velocity signal obtained from the second
sound tweezers anemometer, for two values of $U$ at 1.65 K.
The spectra collapse using the scaling $f/U$ for the frequency
and $P_U(f)/U$ for the spectral density. The straight line displays
the $-5/3$ slope which is expected for a classical velocity spectrum
in the inertial range of the turbulent cascade. The dotted line is a fit using the von K\'arm\'an expression (see \cite{vita2018generating}) to find the integral scale $H$.
\label{fig:spectres}}
\end{figure}

Fig. \ref{fig:spectres} presents the main result of this letter.
We display on the top panel the VLD power spectral density $P_L(f)$ of $L_\perp/\left\langle L_{\perp}\right\rangle$. With this definition, the VLD turbulence intensity $L_{\perp}^{\rm{rms}}/\left\langle L_{\perp} \right\rangle$ is directly given by the integral of $P_L(f)$. We have measured the VLD fluctuations at the temperatures $T=1.65$ K and
superfluid fraction $\rho_{S}/\rho=81\%$, $T=1.99$ K and $\rho_{S}/\rho=47\%$,
$T=2.14$ K and $\rho_{S}/\rho=16\%$. At each temperature, the measurement was done for 
at least two different mean velocities. 

The first striking result is the collapse of all the spectra
independently of the temperature, when properly rescaled
using $f/U$ as coordinate (and $P_L(f)\times U$ as power spectral density to keep the integral constant).
The VLD spectrum does not depend on the superfluid fraction
even for vanishing superfluid fractions, when $T$ comes very close to $T_{\lambda}$. Only one measurement with one of the large tweezers at $T=1.650$ K has given a slight deviation from the master curve of the VLD spectra: it is displayed as the thin grey curve in Fig. \ref{fig:spectres}. We have no explanation for this deviation but did not observe this particular spectrum with the second tweezers, and neither at any other temperature. 

Second, the VLD spectrum has no characteristic
power-law decay. We only observe that the spectrum follows an exponential decay approximately above $f/U>100$ m$^{-1}$. This strongly contrasts with the velocity spectrum obtained with the small second sound tweezers anemometer (see bottom panel),
which displays all the major features expected for a velocity
spectrum in classical turbulence: it has a sharp transition from a plateau at large scale to a power law scaling close
to $-5/3$ in the inertial scales of the turbulent cascade. Actually, it can be seen that the spectral decrease is less steep than $-5/3$, which can be due either to non-perfect isotropy and homogeneity, or more likely because the signal has some second-order corrections in addition to its dependence on velocity fluctuations. A fit of the transition using the von K\'arm\'an expression (see \cite{vita2018generating}) gives the value $H=5$ mm for the longitudinal integral scale. As a side remark, the apparent cut-off above $10^3$ m$^{-1}$ is an instrumental frequency cut-off of the tweezers.

 We find a value of the VLD turbulent intensity close to 20\%,
which is significantly higher than the velocity turbulence intensity. We also checked that we obtain the same VLD spectrum using different
resonant frequencies $f_{0}$. 

Our measurements are limited by two characteristic frequencies. First,
the tweezers average the VLD over a cube of side $l$, which means
that our resolution cannot exceed $f/U>1/l$. For the large tweezers,
this sets a cut-off scale of $10^{3}$ m$^{-1}$, much larger than
the range of inertial scales presented in top panel of Fig. \ref{fig:spectres}.
Second, the frequency bandwidth of the resonator decreases when the
quality factor of the second sound resonance increases. This again sets a cut-off scale given by
$f/U=\xi_{0}c_{2}/(2U)$. The worst configuration corresponds to the
data obtained at 2.14 K and $U=1.2$ m/s where the cut-off scale is about
$600$ m$^{-1}$. For this reason, the VLD spectra of
Fig. \ref{fig:spectres} are conservatively restricted to $f/U<300$ m$^{-1}$ which allows to resolve about one and a half decade of inertial scales.

\begin{figure}
\begin{centering}
\includegraphics[height=6cm]{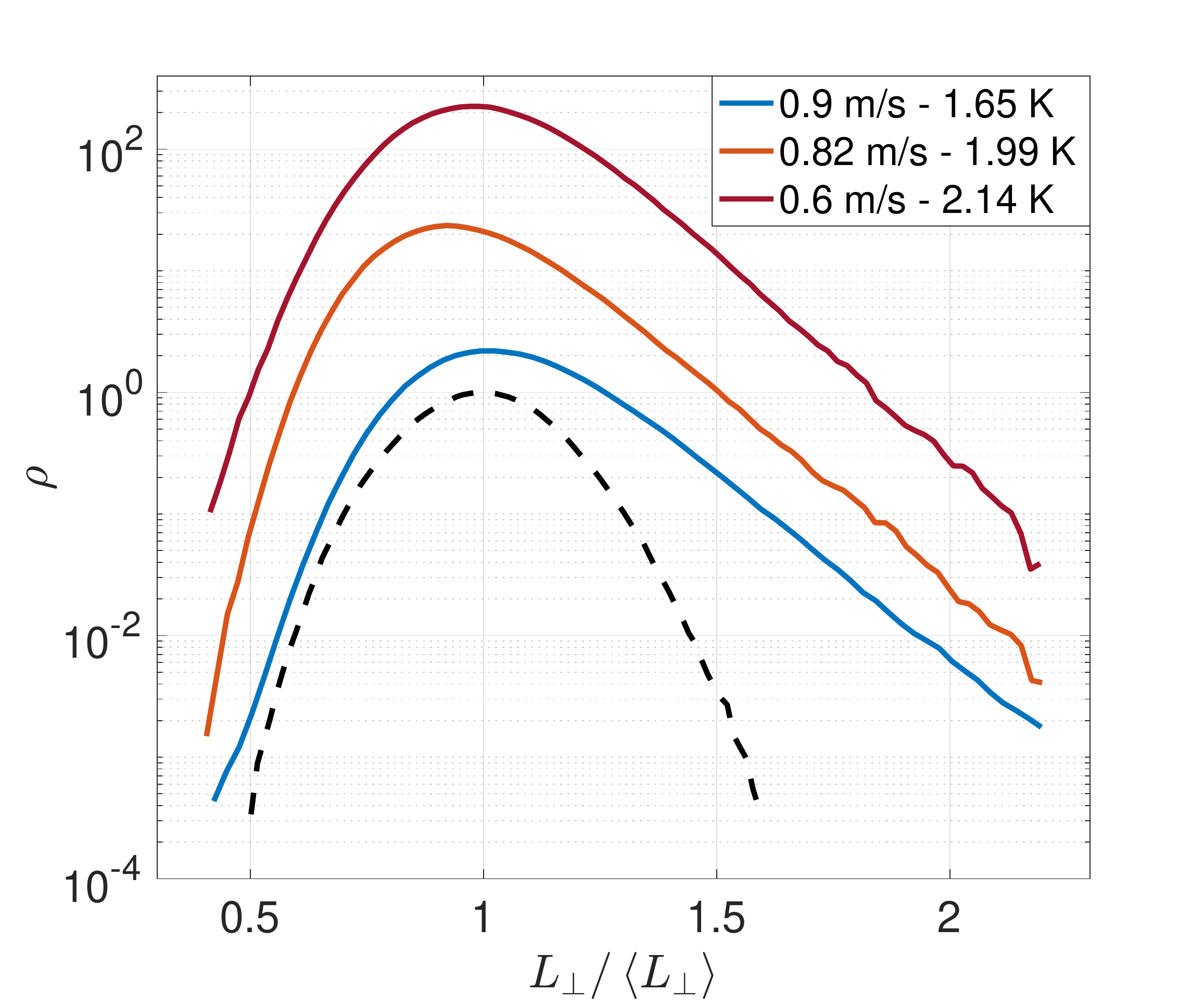}
\par\end{centering}
\caption{Normalized probability distributions of the VLD fluctuations obtained
at three temperatures. The PDF have been shifted by one decade from
each other for readability. By comparison, the dotted black curve displays a rescaled PDF obtained with the small tweezers measuring velocity. \label{fig:histogrammes}}
\end{figure}

Figure \ref{fig:histogrammes} displays some typical PDF of the rescaled
VLD fluctuations $L_{\perp}/\left\langle L_{\perp} \right\rangle$ in semilogarithmic scale, for the three considered
temperatures. The PDF have been vertically shifted by one decade from each other
for readability. The figure shows a strong asymmetry at all temperatures,
with a nearly Gaussian left wing, and an exponential right wing. Contrary to the VLD spectra, the PDF do not accurately collapse on a single master curve at different velocities and temperatures: yet, they remain very similar when the temperature and the mean velocity are changed, and their strongly asymetric shape seems to be a robust feature.
By contrast, the dotted curve in Fig. \ref{fig:histogrammes} displays one PDF of the small tweezers anemometer at $1.65$ K, for which the mean has been shifted and the variance rescaled. It can be seen that the general shape of this latter PDF is much more symmetric and closer to a Gaussian as expected for a PDF of velocity fluctuations.

\section{Discussion and conclusion}

In the present paper, we have investigated the temperature dependence of the statistics of the local density of vortex lines (VLD) in quantum turbulence. About one and a half decade of inertial scales of the turbulent cascade was resolved. We measure the VLD mean value and deduce from Eq. (\ref{eq:tau}) the turbulence intensity (Fig. \ref{fig:scalingU}), we report the VLD power spectrum (Fig. \ref{fig:spectres}), and the VLD probability distribution (Fig. \ref{fig:histogrammes}). Whereas the VLD mean value at different temperatures confirms previous numerical \cite{salort2011mesoscale,Babuin:EPL2014} and experimental studies \cite{Babuin:EPL2014}, the  spectral and PDF studies are completely new. Only one measurement of the VLD fluctuations had been done previously around 1.6K \cite{roche2007vortex} but in a wind tunnel with a very specific geometry and a non-controlled turbulence production. In the present work, we have used a grid turbulence, which is recognized as a reference flow with well-documented turbulence characteristics.

To conclude, we discuss below the three main findings:
\begin{enumerate}
    \item A master curve of the VLD spectra, independent of temperature and mean velocity.
    \item The observed master curve does not correspond to previously reported spectra in the context of highly turbulent classical flows.
    \item A global invariant shape of the strongly skewed PDF.
\end{enumerate}

The mean VLD gives the inter-vortex spacing, and thus tells how many quantum vortices are created in the flow, whereas the PDF and spectra tell how those  vortices are organized in the flow. From 2.14K to 1.65K, our results confirm that the inter-vortex spacing only weakly decreases, by less than 23\% for a 5-times increase of the superfluid fraction. In other words, the superfluid fraction has a limited effect on the creation of quantum vortices. The current understanding of the homogeneous isotropic turbulence in He-II is that the superfluid and normal fluid are locked together at large and intermediate scales where they undergo a classical Kolmogorov cascade \cite{spectra:PNAS2014}. The experimental evidences are based on the observation of classical velocity statistics 
using  anemometers measuring the barycentric velocity of the normal and superfluid components. Here, the temperature-independence of (normalized) VLD spectra supports this general picture, by reminiscence of a similar property of He-II velocity spectra. 

In contrast to velocity, the observed VLD master curve has an unexpected shape in the inertial range, at odd with the spectra reported  as ``compatible with'' a $f^{-5/3}$ scaling in \cite{roche2007vortex}. The probe is sensitive to the total amount of vorticity in the scales smaller than the probe spatial resolution, and thus keeps track of the small scales fluctuations. A close classical counterpart of VLD is enstrophy, because its 1-D spectrum is also related to the velocity spectrum at smaller scales (eg. see \cite{antonia1996note}). However, the experimental \cite{baudet1996spatial} and numerical (e.g. \cite{ishihara2003spectra}) enstrophy spectra reported so far in three-dimensional classical turbulence strongly differ from the present VLD spectra. We have no definite explanation for this difference. It could originate from remanent quantum vortices pinned on the grid, that cause additional energy injection in the inertial range, in which case the peculiarity of our spectra would be specific to the type of forcing. Otherwise, it could be a more fundamental property associated with the microscopic structure of the vortex tangle that, together with the observed temperature-independence of the spectra, would be very constraining to develop mathematical closures for the continuous description of He-II (eg. see \cite{nemirovskii2020closure}).

As a discussion of the third statement,
we compare the PDF with those of numerical simulations done in classical turbulence. The absolute value of vorticity can be seen as a classical counterpart to the VLD. The work of Iyer and co. \cite{yeung2015extreme} for example, displays some enstrophy PDF from high resolution DNS, that can be compared to the PDF of Fig. \ref{fig:histogrammes}.
At small scale, the enstrophy PDF are strongly asymmetric and will ultimately converge to a Gaussian distribution when averaged over larger and larger scales. Although our tweezers average the VLD over a size much larger than the inter-vortex spacing, they are small enough to sense short-life intense vortical events, typical of small scale phenomenology in classical turbulence. Thus, the strong asymmetry of the PDF supports the analogy between VLD and enstrophy (or its square root) and shows the relevance of VLD statistics to explore the small scales of quantum turbulence.

A side result of the present work is to obtain the relative values of
the empirical coefficient $\nu_\mathrm{eff}=\epsilon(\kappa \mathcal{L})^{-2}$ at the three considered temperatures. Models and simulations predict that $\nu_\mathrm{eff}$ should steeply increase close to $T_\lambda$ (see \cite{Babuin:EPL2014,boue2015energyVorticity,gao2018dissipation} and ref. within), 
in apparent contradiction with the only  systematic experimental exploration \cite{stalp2002}. We found in Fig. \ref{fig:scalingU} that the effective viscosity $\nu_\mathrm{eff}$ is twice larger at 2.14K than at 1.99K.
To the best of our knowledge, our estimate $\nu_\mathrm{eff}(2.14 K) \simeq 2\,(\pm 0.25)\times \nu_\mathrm{eff}(1.99K)$ is the first experimental hint of such an effective viscosity increase.

\acknowledgments
We warmly  thank B. Chabaud for support in upgrading the wind-tunnel and P. Diribarne, E. Lévêque and B. Hébral for their comments.
We thank K. Iyer with his co-authors for sharing data on the statistics of spatially averaged enstrophy analyzed in \cite{iyerNJP2019}.
Financial support from grants ANR-16-CE30-0016 (Ecouturb) and ANR-18-CE46-0013 (QUTE-HPC).
\bibliographystyle{eplbib}
\bibliography{mabiblio}
\end{document}